# On the Early History of Current Algebra

by

## Herbert Pietschmann

## Univ. of Vienna, Faculty of Physics

Abstract:
The history of Current Algebra is reviewed up to the appearance of the Adler-Weisberger sum rule. Particular emphasis is given to the role current algebra played for the historical struggle in strong interaction physics of elementary particles between the S-matrix approach based on dispersion relations and field theory. The question whether there are fundamental particles or all hadrons are bound or resonant states of one another played an important role in this struggle and is thus also regarded.

## 1. S-Matrix versus Quantum Field Theory in the 1960's

Right after the end of World War 2, particle physics received its relativistic and mathematically sound basis by the new renormalization theory; among many contributors, Feynman, Schwinger and Tomonaga were distinguished by the Nobel Prize of 1965 "for their fundamental work in quantum electrodynamics, with deep-ploughing consequences for the physics of elementary particles."

For the first time, it was possible to calculate higher-order corrections ("loop diagrams"), the most famous examples were the anomalous magnetic moment of electrons and muons, and the Lamb shift.

However, this great victory in Quantum Electrodynamics was overshadowed by the fact that it did not work as well for strong interactions and not at all for weak interactions (except for the lowest order, which worked astonishingly well).

Thus it became necessary to look for a different approach. In 1956, at the $6^{th}$ annual Rochester conference taking place in New York, M. Gell-Mann suggested to use dispersion relations to calculate observable quantities.

Two years later, S. Mandelstam published a historic paper on double dispersion relations [1]; (soon they became known as the "Mandelstam representation"). In the introduction, Mandelstam wrote:

"In recent years dispersion relations have been used to an increasing extent in pion physics for phenomenological and semiphenomenological analyses of experimental data, and even for the calculation of certain quantities in terms of the pion-nucleon scattering amplitude. It is therefore tempting to ask the question whether or not the dispersion relations can actually replace the more usual equations of field theory and be used to calculate all observable quantities in terms of a finite number of coupling constants."

The idea to *replace* field theory was taken up by a number of theoreticians who did not like the concept of renormalization, for it involves the notion of unobservable hence unphysical "bare" particles and the renormalization procedure which used mathematically ill-defined quantities. From the point of view of positivism, the notion of "relativistic field", which is not directly observable, was also of concern for them.

Most vigorously, Geoffrey F. Chew fought for a new approach. He wanted to base particle theory on observable quantities only. S-Matrix (scattering matrix) elements should replace quantum fields with the postulate [2]: "*The S matrix is a Lorentz-invariant analytic function of all momentum variables with only those singularities required by unitarity.*"

Quantum Field Theory was originally designed to describe electromagnetic interactions, i.e. the interaction of photons with electrons and muons. Strongly interacting particles, the hadrons, formed a much larger sample. At the time, eight fermions ($p, n, \Lambda, \Sigma^+, \Sigma^0, \Sigma^-, \Xi^0, \Xi^-$) plus their anti-particles and seven bosons ($\pi^0, \pi^+, \pi^-, K^0_1, K^0_2, K^+, K^-$) were known. The question came up, whether some of them were truly elementary with the others just bound states of the former.

Chew had in mind an even more progressive change in attitude. He wanted to develop a theory in which the difference between elementary and composite particles would disappear: "It is difficult, however, to imagine a calculation sufficiently complete to approach a definite answer to the question: *Which* of the strongly interacting particles are elementary? Partly because of this circumstance, but even more because of general philosophical conviction, I am convinced that there can be only one sensible answer, and that is that *none* of them is elementary. … In particular there is a remark often made privately by Feynman that tends to convert the negative statement into a positive one. Paraphrasing Feynman: *The correct theory should be such that it does not allow one to say which particles are elementary.* Such a concept is manifestly at odds with the spirit of conventional field theory, but it forms a smooth alliance with the S-matrix approach."

Chew's approach soon became known as the "bootstrap-mechanism" and – for some time – was taken up by quite a number of theorists. Some went so far as to request that quantum field theory should be abandoned from the curricular of universities. B. Schroer describes this struggle with the following words [3]: "This led to a confrontation of the S-matrix bootstrap with Quantum Field Theory at the end of the 60s. It was a struggle about a pure S-matrix approach cleansed of all field theoretic aspects; … The ideological fervor found its strongest expression in conference reports were the S-matrix bootstrap proponents felt more free to celebrate what they perceived as their (premature) victory over Quantum Field Theory. … The ferocity of the struggle on the side of the S-matrix purist against Quantum Field Theory is hard to understand in retrospect, … ".

Let us recall that Quantum Field Theory starts from so-called "bare" particles without interaction. (They carry "bare" masses and coupling constants). When the interaction is turned on, these quantities change into "physical" masses and coupling constants. The difference can – in principle – be calculated, but the relevant integrals do not converge, i.e. they turn out to be infinite (mathematically speaking: they do not exist!). In so-called "renormalizable theories" (including quantum electrodynamics!) observable quantities do not depend on these infinite integrals. Thus by a largely arbitrary "regularization", the integrals can be forced to converge, because they do not enter in the observables. Although – as mentioned – results were beautifully compatible with experiments (e.g. the magnetic moment of electrons and muons), the procedure left some uneasiness even with its advocates: "Infinities are swept under the rug!" was often bemoaned.

Let me quote extensively from Chew in order to shed some light on the situation [2]: "So that there can be no misunderstanding … let me say at once that I believe the conventional association of fields with strongly interacting particles to be empty. I do not have firm convictions about leptons or photons[1], but it

---

[1] Consequently, the physics of leptons and photons as well as of weak interactions found itself in a marginal position at large conferences. This led to the creation of new types of conferences. 1962 saw the first of a now well established series, the "Conference on electron-photon interactions in the BeV energy range" in Boston, now called "International Lepton-Photon Symposium". G. Marx created the series of

seems to me that no aspect of strong interactions has been clarified by the field concept. … I do not wish to assert … that conventional field theory is necessarily wrong, but only that it is sterile with respect to strong interactions and that, like an old soldier, it is destined not to die but just to fade away. Having made this point so strongly, I hasten to express an unqualified appreciation of the historical role played by field theory up to the present. … However, it is my impression now that … future development of an understanding of strong interactions will be expedited if we eliminate from our thinking such field-theoretical notions as Lagrangians, 'bare' masses, 'bare' coupling constants, and even the notion of 'elementary particles'. I believe, in other words, that in the future we should work entirely within the framework of the analytically continued S matrix."

But already in the preface Chew concedes: "Readers should be aware that S-matrix theory is still incomplete … . Progress is currently rapid, and a complete theory may well develop within a few years' time. (A paper by H. Stapp, currently in press, makes a major step in this direction.)"

The mathematically most rigorous results of field theory were the CPT theorem and the connection between spin and statistics. With very few assumptions, invariance under the combined charge conjugation (C), parity (P) and time-reversal (T) was proven, though neither of the three alone is conserved. Likewise it was proven that particles with integer spin (bosons) follow Bose-Einstein statistics and particles with half integer spin (fermions) obey Fermi-Dirac statistics.[2]

In the abstract of the mentioned paper H. Stapp writes [4]: "The CPT theorem and the normal connection between spin and statistics are shown to be consequences of postulates of the *S*-matrix approach to elementary particle

---

conferences on neutrino physics and astrophysics, the first of which took place at Lake Balaton in Hungary 1972 and the next year saw the first "Workshop on Weak Interactions with Very High Energy Beams, created by Jan Nilsson and the author in Skövde, Sweden, now called Workshop on Weak Interactions and Neutrinos (WIN).

[2] A comprehensive description can be found in R. F. Streater and A. S. Wightman: PCT, Spin and Statistics, and All That. Reprinted by Addison-Wesley, New York, (1989).

physics. The postulates are much weaker than those of field theory. Neither local fields nor any reference to space-time points are used. Quantum commutation relations and properties of the vacuum play no role. Completeness of the asymptotic states and positive definiteness of the metric are not required, though certain weaker asymptotic conditions prevail. The proofs depend on unitarity, macroscopic relativistic invariance, and a very weak analyticity requirement on the mass-shell scattering functions. The proofs are in the framework of the new *S*-matrix approach to elementary particle physics, which is established on a formal basis."

Stapp concedes that the CPT Theorem and the connection between spin and statistics are "the two most important general physical consequences of relativistic field theory" but he objects to field theory for "it is not known whether the postulates are sufficiently realistic to include any theories except trivial ones in which the scattering matrix is unity." Moreover, from a philosophical point of view he insists that "experience does not entail the existence of space-time points" which are at the basis of relativistic quantum field theory. He insists: "Because space-time points are experimentally inaccessible both in practice and in principle, their introduction runs counter to the philosophy of quantum mechanics."

On the other hand, according to Stapp, "S-matrix approach to elementary particle physics is … approaching the status of an independent theory, its connections to field theory gradually being dissolved. … History encourages the casting away of formal substructures whose ingredients have no counterparts in experience … . The new approach, since it involves only observable quantities and their analytic continuations, has a claim to probable physical relevance much greater than that of field theory, with its sundry hypothetical ingredients of dubious status."

Concluding, Stapp writes: "It has been shown how the appeal to field theoretic concepts can be completely avoided and the new S-matrix formalism built up from simple principles that are relatively secure."

Obviously, this far reaching claim met fierce opposition from field theory adherents. One of the most outspoken opponents was Res Jost [5]. He wrote: "The main vice of Mr. Stapp is the fact that he does not properly reflect upon the problems."[3] and "It seems that Mr. Stapp does not shy away from arbitrarily criticising other physicists, but for his own case he is missing those self criticism and precision of formulating which one is used to require from a physics paper."[4]

This fierce controversy on the philosophical side was paralleled from the experimental side by the unexpected discovery of a multitude of strongly interacting particles (hadrons).

## 2. Truly elementary particles or "particle democracy"?

In November 1959, the Proton Synchrotron at CERN began to operate at an energy of 28 GeV. A few months later, a similar machine – the Alternating Gradient Synchrotron at Brookhaven National Laboratory – reached an energy of 33 GeV.

Soon after their inauguration, the new machines produced an unexpectedly large number of short-lived hadrons, then called "resonances". Although their life-times of about $10^{-22}$ sec were very short, there was little principal difference to some of the "stable" particles, e.g. the neutral pion with a life-time of $10^{-16}$ sec.

---

[3] German original: Das Hauptübel besteht bei Herrn Stapp eben darin, dass er sich die Probleme nicht eigentlich überlegt hat …

[4] German original: Uns scheint, dass Herr Stapp, der es anderen Physikern gegenüber an wahlloser Kritik nicht fehlen lässt, in eigener Sache auch diejenige Selbstkritik und Schärfe der Formulierung vermissen lässt, die man von einer physikalischen Arbeit zu verlangen gewohnt ist …

This avalanche of new "elementary" particles further enhanced the contradiction between the idea of "bootstrap" and field theory based on truly elementary entities.[5]

In the midst of this controversy, a totally new idea was created by M. Gell-Mann [8] and G. Zweig [9]: Neither should there be *no* elementary particle, nor should some out of the "zoo" be *more elementary* than the others; rather, a totally new class of particles – unobserved at that time – should be the constituents of all the known "elementary particles". These truly fundamental entities were called "quarks" by Gell-Mann (with reference to Finnegans Wake by James Joyce) and – inspired by card games – "aces" by Zweig. Gell-Mann left it open, whether they were actual physical particles or "mathematical entities": "It is fun to speculate about the way quarks would behave if they were physical particles of finite mass, instead of purely mathematical entities as they would be in the limit of infinite mass."

This idea was so out of the conventional thinking of the time, that in the beginning, most of the physicists simply did not take it seriously. After all, the charges of these "fundamental particles" were predicted to be $2e/3$ and $-e/3$ ($e$ = fundamental charge). This seemed to be in contradiction with the classic experiments of Millikan from 1911.

Consequently, Zweig could not publish his paper and Gell-Mann sent his paper to the relatively new "Physics Letters"; before elaborating on his alternative idea, Gell-Mann concedes: "A highly promising approach is the purely dynamical 'bootstrap' model for all the strongly interacting particles within which one may try to derive … broken eightfold symmetry from self-consistency alone."

The astonishment of the majority of physicists is best disclosed by H. Lipkin, who sent a paper to Physics Letters which was originally intended as a joke [10]. (The original title was "The barbaryon classification for elementary particles

---

[5] First individual lists of the new states were published in 1963 [6]. From 1964, a group of physicists started to publish annual reviews of particle properties [7], since 1969 it is officially quoted as "Particle Data Group".

SU(3)".) He writes: "Since possibilities are now being considered which seem just as fantastic as the barbaryon classification, perhaps the latter is not so crazy after all and deserves more serious consideration."

The astonishment as well as excitement of the physics community is best recalled by the issue of CERN Courier from March 1964:

"From the time of Millikan's classic experiments in 1911 it has been accepted that the charge of the electron is the smallest one possible. The idea of fractionally charged particles seemed quite preposterous. Even those who suggested it seemed to share the doubts; … For experimentalists, the excitement lay in the prediction that at least one of the new particles would be stable. … The Electronics Experiments Committee, on 11 February, decided that the particles should be taken seriously. … In any case, it quickly became clear that the combination of a bubble chamber and the $o_2$ beam in the PS East hall provided the quickest way of looking for them. … While working on this proposal, D R O Morrison realized that the same kind of bubble-chamber exposure had in fact been carried out with the CERN 32 cm chamber in 1960. The photographs were got out and a team of physicists and scanners looked through 10.000 of them in one night. No aces were found. The group working with the Ecole Polytechnique heavy-liquid bubble chamber scanned 100.000 photographs. Again the result was negative. … Zweig's aces and Gell-Mann's quarks may or may not be found, but their ideas have triggered off a new series of moves in this search for an explanation of the occurrence of the so-called fundamental particles."

Particles with fractional charge were indeed found in cosmic ray events! McCusker and Cairns write [11]: "In one year from July 1968 we found four tracks whose appearance was that expected for a quark of charge 2e/3." But the result could not be reproduced.

About ten years later, free quarks turned up again. A Millikan-type experiment showed fractional charges on Niobium balls [12]. It was not sufficient objection that no such effects were found in similar experiments with steel balls [13]. After all, quarks could have a preference for certain elements. In order to rule out this possibility, the experiment had to be repeated with Niobium [14]. This is a fine example for the author's claim [15]: "Everything that is predicted by a sufficiently renowned theorist will be discovered, irrespective of its actual existence; that is why in physics the criterion for existence is not a discovery, but only the proof of reproducibility!"

From these discussions, it was remembered that even at the time of Millikan there was a great controversy [16]. Felix Ehrenhaft at the University of Vienna who had independently invented the method used by Millikan, kept producing fractional charges. Thus on June 11, 1980 G. Zweig wrote the following letter to the author:

"Two students of mine are repeating the Millikan oil drop experiment using a number of materials, including selenium, instead of oil. In Johanna Fürst's 1920 Dissertation from the Third Physics Institute of the University of Vienna, the measured charges an 150 selenium spheres are given. These results, which are striking to the modern eye, are published by Felix Ehrenhaft in Physik.Zeitschr. 39, 673 (1938). … Please note the two large peaks at charge 1 and 2/3. There are so many possible explanations for the peak at 2/3. For example, if the selenium was in the form of two spheres sticking together and had a net charge of 1, then the apparent charge coming out of Fürst's analysis would have been less than one. Nevertheless, the possibility, however remote, that free quarks were present in the selenium is still there. Consequently we would like to have a sample of the selenium used by Fürst in 1920 (and perhaps by Ehrenhaft at a later time). As a favour, would you please find out if this selenium still exists? If so, please send a sample. If not, is it possible to find out who manufactured this selenium? If all else fails, would it be possible to get some selenium that was sold in Vienna before 1920? Perhaps there is an old bottle stored away in one of the chemical stockrooms.

Thank you for your assistance. Sincerely yours George Zweig."

Unfortunately I could not be of any help. Johanna Reif-Fürst had passed away and among many materials from that time just selenium was neither left over nor was its source traceable. Two former assistants of Ehrenhaft had become Professors of physics, but their search did not produce any result either. Eventually, quarks were discovered in a reproducible way, however neither as free particles nor as mathematical entities, rather confined within the nucleons. The Nobel prize in physics was given in 1990 to J.I. Friedman, H.W. Kendall and R.E. Taylor "for their pioneering investigations concerning deep inelastic scattering of electrons on protons and bound neutrons, which have been of essential importance for the development of the quark model in particle physics." Quantum-Chromodynamics now provides the theory capable of predicting the phenomenon of confinement.

In a certain sense, history seems to repeat itself periodically on ever deeper levels of understanding: In its October 2010 issue (p.6), the CERN Courier writes: "The ATLAS experiment at the LHC has set the world's best known limits for the mass of a hypothetical excited quark, $q^*$. … The existence of such a state would indicate that a quark is a composite particle as opposed to an elementary one as the Standard Model assumes."

### 3. An Observable Renormalization

Let us turn back to the controversy between S-Matrix and Field Theory approaches. If it were possible to construct a model which can only be described by unitarity and dispersion relations, but does not follow from a Lagrangian (or Hamiltonian), it would mean victory for the S-Matrix approach.

In 1961, F. Zachariasen claimed exactly that breakthrough [17]. He writes: "We shall construct a model field theory which is perhaps unusual in that it is not defined in terms of a Lagrangian or Hamiltonian; … Instead we shall assume the existence of a complete set of dispersion relations, which, together with unitarity, form a set of coupled integral equations for the transition amplitudes of the theory."

And, more explicitly: "Conventionally, field theories are defined by specifying a Lagrangian density, and from this obtaining field equations, perturbation expansions, and so on. It has been suggested,[6] however, that field theories can equally well be defined by writing down a set of dispersion relations which, when combined with unitarity, provide an infinite set of coupled integral equations from which all the transition amplitudes of the theory may be determined. This second approach has the virtue of not involving any unobservable quantities, such as bare masses or coupling constants, at any stage of the development."

Within the same year, W. Thirring published a paper called "Lagrangian Formulation of the Zachariasen Model" [18]. He writes: "Our results have shown that there exists a Lagrangian for the Zachariasen model which has the same short-comings as the one of other relativistic field theories: It contains renormalization constants when expressed in terms of bare fields but everything is perfectly finite when the incoming or outcoming fields are introduced. … Our findings make one suspect that field theories without Hamiltonian, which are defined only by dispersion relations, actually have an underlying Hamiltonian structure."

We have seen that the defenders of S-Matrix approach refer to M. Gell-Mann's 1956 remark at the Rochester Conference in New York. But in his fundamental paper [19] from 1962 he cautiously turns towards field theory. In this paper, Gell-Mann considers the algebraic structure of the electric current $j_\alpha$ and the weak current $J_\alpha$. Already in the abstract, he notes that matrix elements of these currents are well defined and obey dispersion relations; however, "homogeneous linear dispersion relations, even without subtractions, do not suffice to fix the scale of these matrix elements; in particular, for the nonconserved currents, the renormalization factors cannot be calculated,...". In other words, the important ratio of weak axial over weak vector nucleon current, $G_A/G_V$, cannot be calculated by linear relations. Since it is an ingredient in many fundamental relations (such as the Goldberger-Treiman-Relation), its theoretical computation was considered an important task. Thus Gell-Mann continues in the abstract: "More information than just the dispersion relations must be supplied, for example, by field-theoretic models; we consider, in fact, the equal-time commutation relations of the various parts of $j_4$ and $J_4$.[7] These nonlinear relations define an algebraic system (or a group) that underlies the structure of baryons and mesons."

---

[6] Zachariasen refers to the talk of M. Gell-Mann at the Rochester Conference in New York 1956, see chapter 1.

An internal symmetry is defined by the algebra of its generators

$$[I_\alpha, I_\beta] = c_{\alpha\beta\gamma} I_\gamma \qquad (1)$$

(In the case of Isospin, the $c_{\alpha\beta\gamma}$ are the components of the ε-Tensor times $i$.)
The generators, in turn, are given by the integral over the time-component of the currents

$$I_\alpha = \int d^3 x J_{0,\alpha}(x) \qquad (2)$$

From these equations one obtains the equal-time commutation relation of the currents

$$\left[ J_{0,\alpha}(\vec{x}), J_{0,\beta}(\vec{y}) \right] = c_{\alpha\beta\gamma} J_{0,\gamma}(\vec{x}) \delta(\vec{x}-\vec{y}) \qquad (3)$$

Since these are nonlinear equations, they allow for a computation of $G_A$ which is the matrix element of the axial-vector current between proton and neutron states at zero momentum transfer.

---

[7] the time-components of the currents, now usually denoted by $j_0$ and $J_0$.

In the text, Gell-Mann points this out explicitly: "The dispersion relations for the matrix elements of weak or electromagnetic currents are linear and homogeneous. … Now such linear and homogeneous equations … cannot fix the scale of these matrix elements; constants like $-G_A/G$ cannot be calculated without further information. A field theory of the strong interactions, with explicit expressions for the currents, somehow contains more than these dispersion relations."

But in the final paragraph he insists: "Nowhere does our work conflict with the program of Chew *et al.* of dynamical calculation of the *S* matrix for strong interactions, using dispersion relations. … If there are no fundamental fields … , all baryons and mesons being bound or resonant states of one another, … the symmetry properties that we have abstracted can still be correct."

In a sense, the axial-vector constant $G_A$ is an observable renormalization constant and as such weakens the objection of S-Matrix defenders against field theory because of its unobservable renormalization.

Gell-Mann suggests that "the equal-time commutation relations for currents and densities lead to exact sum rules for the weak and electromagnetic matrix elements." This suggestion eventually led to the technique of current algebra and yielded the long awaited computation of $G_A/G_V$, in Gell-Mann's notation $-G_A/G$.

**4. First Attempts**

In the fall of 1962, A.P.Balachandran from the Institute of Mathematical Sciences in Madras, India, came to Vienna as postdoc in Walter Thirring's research group. He had just studied the paper by Murray Gell-Mann and – together with the author – wanted to take up the above mentioned suggestion. Since the equal-time commutation relations of the weak isospin currents are not explicitly covariant, the choice of frame is imperative. We found the rest frame to be the most natural choice. However, we soon realized that we could not get rid of infinities when non-conserved currents were involved. Thus we could not arrive at our original goal, the calculation of $G_A/G_V$. So we sadly turned to the algebra of electromagnetic iso-vector currents in order to work out the method of current algebra [20]. We could not go beyond the lowest (one-nucleon intermediate state) approximation: "Our concern here is more with the fact that even in the lowest approximation the equal time commutation relations of the isospin density imply certain non-trivial restrictions on the structure of the isovector form factors and once these implications are analysed with adequate accuracy, it will no longer be permissible to ignore them even in a dispersion theory treatment, since such a neglect would mean a violation of isospin invariance, albeit in a rather subtle fashion." (The next approximation was carried out in a subsequent paper [21].)

A current algebra calculation in the static model was presented at the 1963 Siena International Conference on Elementary Particles [22]; it prompted M. Gell-Mann in his summary talk at the conference to encourage further work in current algebra.

On the occasion of a visit at the Univ. of Vienna's Institute for Theoretical Physics, Sergio Fubini learned about the new method. He realized that our problems with infinities could be overcome, if instead of the rest frame, an infinite-momentum frame is used. In a paper together with G. Furlan they write [23]: "Thus on the basis of this model we are led to the conclusion that the quantity $\varphi(p)$ is a minimum when $p \to \infty$, and according to our previous discussion this fixes the 'best' sum rule". But they did not achieve to calculate $G_A/G_V$ either. One small step was still missing, filled in subsequently by S. Adler and W. Weisberger.

## 5. The Breakthrough

It is a well-known fact in history, that important developments are often initiated simultaneously but independently by more than one person. Such was the case with the calculation of the renormalization of the weak axial-vector coupling constant: In the same issue of Physical Review Letters, Stephen Adler [24] and William Weisberger [25] published their version of the calculation of $G_A/G_V$. (The result became known as the "Adler-Weisberger sum rule".)

Both calculations are based on the paper by Fubini and Furlan [23] which was still unpublished at that time; the missing step filled in by Adler and Weisberger was the assumption that the axial-vector current is "partially conserved". That means, that the divergence of the weak axial-vector current is neither zero nor unknown, but is proportional to the pion field $\varphi(x)$.

$$\frac{\partial}{\partial x_\mu} J_\mu(x) = \frac{\sqrt{2} M \mu^2 G_A}{g} \varphi(x) \qquad (4)$$

where $M$ is the nucleon mass, $\mu$ the pion mass and $g$ the pion-nucleon coupling constant.

This hypothesis had been suggested by Gell-Mann and Lévy [26] and further developed by several authors [27]. The result was a "sum rule expressing the axial-vector coupling-constant renormalization in $\beta$ decay in terms of off-mass-shell pion-proton total cross sections." (S. Adler). It was in good agreement with experiment.

The Adler-Weisberger sum rule was a breakthrough which settled the method of current algebra as a widely applicable tool and inspired others to take up this approach. Steven Weinberg became one of the most successful researchers in this field. (The author lectured on current algebra in many places [28]).

This further development is very well described in a paper by Steven Weinberg [29]; thus we can leave the description of the development of current algebra at that historical point and refer the reader for the following events to the paper of Steven Weinberg.

Acknowledgement: The author thanks Gerhard Ecker for a critical reading of the manuscript.